# Giant Nonlinear Response of 2D Materials Induced by Optimal Field-Enhancement Gain Mode in Hyperbolic Meta-Structure

Hao-Fei Xu[1], Ying Yu[1], Limin Lin[1], Zhang-Kai Zhou[1,*], Xue-Hua Wang[1,*]

[1]State Key Laboratory of Optoelectronic Materials and Technologies, School of Physics, Sun Yat-sen University, Guangzhou 510275, China

Resonant modes in metamaterials have been widely utilized to amplify the optical response of 2D materials for practical device applications. However, the high loss at the resonant mode severely hinders metamaterial applications. Here, we introduce a field-enhancement gain (FEG) factor to find the FEG mode for significantly improving light-matter interaction. As a demonstration, we experimentally compared the second harmonic generation enhancement of monolayer $MoS_2$ induced by the optimal FEG and resonant modes in hyperbolic meta-structures. With the optimal FEG mode, we obtained an enhancement of 22145-fold and a conversion efficiency of $1.1 \times 10^{-6}$ $W^{-1}$, which are respectively one and two orders of magnitude higher than those in previous reports of monolayer $MoS_2$. A broadband high-FEG region over ~80 nm where the nonlinear enhancement is larger than that induced by the resonant mode is achieved. The concept of FEG factor is general to metamaterials, opening a new way for advancing their applications.

The past decades have witnessed remarkable triumphs and revolutionary advances of two-dimensional layered materials (2DLMs) due to their ultrathin thickness. For example, the electronic energy bands of the 2DLMs are reshaped and various quantum confinement effects have been observed [1-3]. Because of the inversion asymmetry, the 2DLMs of odd few-layer (such as the MoS$_2$, h-BN, GaSe, etc.) possess large second-order nonlinear susceptibility with only atomic thickness [4]. However, also due to the ultrathin thickness of 2DLMs, their optical signals are usually very weak, which seriously limits their practical applications.

To address this limitation, the resonant modes of meta-structures are widely used to enhanced the light-matter interaction in 2DLM system and therefore amplify the optical response of the 2DLMs [5-16]. This is because the resonant modes have an extraordinary capacity of highly confining electromagnetic fields (EMFs) to volumes smaller than the diffraction limit, which leads to huge enhancements of EMFs and large increasing of localized photon density of state (LDOS). The resonant modes are identified by the maximum values of the field-enhancement (FE) factor $f_0(\omega)$ ($f_0(\omega) = E(\omega)/E_0(\omega)$, where the $E_0(\omega)$ and $E(\omega)$ respectively represents the amplitudes of incident and total light fields). Therefore, the resonant modes of the metamaterials have long believed to be the best choice for enhancing the light-matter interactions.

However, it is well-known that the resonant modes are always accompanied by large light absorption-loss, which can dramatically reduce actual number of photons involved in light-matter interactions and then significantly decreases actual optical enhancements, severely limiting the applications of the metamaterials [17-20]. A major concern naturally arises whether the resonance modes achieve the maximum enhancement of light-matter interaction. Is there any way to find the balance between the EMF enhancement and the absorption-loss for getting the optimal optical response gain? To address the issues, we introduce the FEG factor $f_{\text{FEG}}(\omega) = [f_0(\omega)]^2 [1-\beta(\omega)]$ that is proportional to the actual number of photons involved in light-matter interactions, where $\beta(\omega)$ is the absorption ratio of the metamaterials at frequency $\omega$. Correspondingly, we define the mode with the largest $f_{FEG}(\omega)$ value as the optimal FEG mode, and the high-FEG region where the FEG factor is larger than that at the resonant mode. In the following text, we will theoretically and experimentally demonstrate that the FEG factor $f_{FEG}(\omega)$ is a more universal physics quantity than the FE factor $f_0(\omega)$ for describing optical enhancement response of the metamaterials, and the optimal FEG mode can lead to much stronger nonlinear enhancement of monolayer $MoS_2$ than the resonant modes.

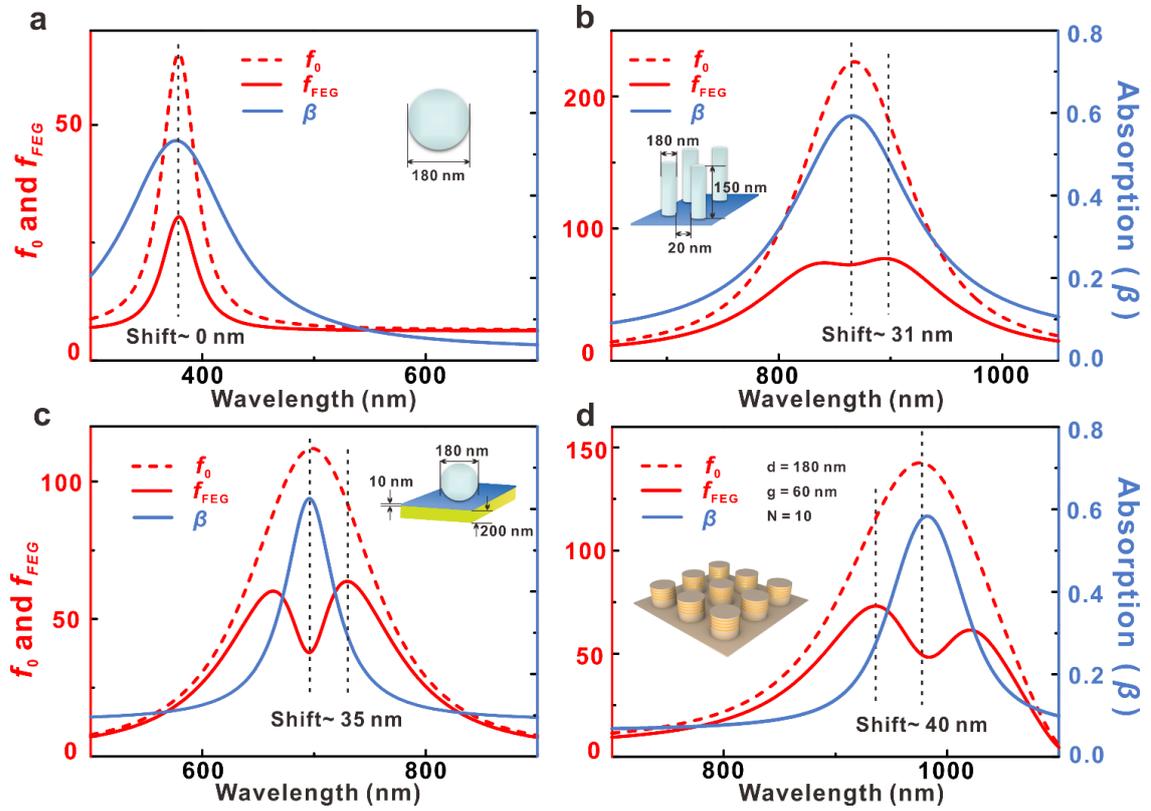

FIG. 1. The comparison of $f_{FEG}$ and $f_0$ factors when they are applied for describing optical enhancement. (a)-(d) Calculated $f_0$, $\beta$ and $f_{FEG}$ curves of some typical nanostructures, including Ag sphere nanoparticle, Ag nanopillar array, Ag nanoparticle on the dielectric/Au substrate, and the periodic hyperbolic metamaterials. The structural parameters of four megastructures are selected based on typical values, and they are marked correspondingly. The refractive index of dielectric layer is 3.2. For the hyperbolic metamaterial, the gap distance g between adjacent multilayered nanopillars is 40 nm, the diameter d of an individual nanopillar is 180 nm, and the number N of layer

is 10 (the definitions of g, d, and N are also marked in Fig. 2a).

Our studies began with theoretical comparisons of the FEG factor $f_{FEG}$ and FE factor $f_0$ for four different nanostructures (Fig. 1). It is clearly seen that the resonant modes are usually accompanied by the largest absorptions. The maximums of the $f_0$ and $f_{FEG}$ occur at the same wavelength only for a single nanosphere structure (Fig. 1a), and appear at different wavelengths for other three meta-structures including the metal nanorods array, metal nanoparticle on film and hyperbolic metamaterial (HMM) (Fig. 1b-d). These results indicate that the optimal FEG mode may be the resonant mode, but the non-resonant mode in the most cases. More interestingly, there is a broadband high-FEG region in which the non-resonant modes have larger FEG factor than the resonant mode. Since the $f_{FEG}$ factor reflects the actual number of photons involved in light-matter interactions, the non-resonant modes in the high-FEG region can significantly improve the light-matter interaction compared with the resonant mode, which is quite different from the traditional concept that the resonant modes are the best for light-matter interaction in metamaterial systems.

Next, we experimentally demonstrate that, comparing with the resonant modes, the non-resonant modes with larger $f_{FEG}$ can significantly improve nonlinear response enhancement of monolayer $MoS_2$ when it was interacted with the hyperbolic metamaterials (HMM). The monolayer $MoS_2$ has exhibited numerous unprecedented properties and potential applications, so using metamaterials to further amplify its optical response is of great importance [21,22]. Owing to their specific structure feature of two-dimensional layer, enhancing the optical response of the 2DLMs requires big EMF enhancement on the metamaterial surface. To meet this requirement, we intentionally fabricated the periodic HMM (PHMM) which is an array of multilayered nanopillars (Fig. 2a). It is well-known that the uniform HMMs can support large EMF enhancement, but their photonic modes are unable to be excited by external light, making them impossible to be directly applied to surface optical enhancements [23,24]. Nevertheless, for the PHMM, the Bragg scattering of periodic multilayered nanopillars can add extra wave vector for the external light, which compensates the momentum mismatch between the wave vectors of external and internal. As a result, unlike the uniform HMM whose reflection spectrum shows a high reflectance approaching 1 in the hyperbolic region (**discussed in Fig. S1**), the reflection curve of the PHMM exhibits an obvious spectral valley (Fig. 2b i), showing successful excitation in PHMM (*i.e.* the high-k mode). Figure 2b ii displays the electric field distributions of a PHMM under excitations of three different wavelengths, demonstrating strong surface EMF enhancement in a broadband spectral region. This advantage can be attributed to the hyperbolic dispersion and Bragg scattering features of the PHMM [23], which is discussed in **Fig. S2**.

In order to maximally improve the nonlinear response enhancement of the monolayer $MoS_2$, structural optimizations of the PHMM have been compressively performed (detailed results are given in **Fig. S3**). Generally, the optimal FEG mode with the largest $f_{FEG}$ value and the bandwidth of the high-FEG region can be manipulated by the gap distance g between adjacent multilayered nanopillars, the diameter d of an individual nanopillar, and the number N of layer (marked in Fig. 2a). In Fig. 2c, we present the calculated $f_0$, $\beta$ and $f_{FEG}$ curves of an optimized PHMM with g = 10 nm, d = 180 nm, N = 10. For this PHMM, the $f_{FEG}$ value of its optimal FEG mode (i.e. the maximum $f_{FEG}$) is about 2 times of the $f_{FEG}$ value giving by the resonant mode, providing the possibility for a notable improvement of nonlinear optical enhancement. Also, a high-FEG region as wide as ~350 nm has been found, suggesting a broadband nonlinear enhancement. In the investigation of structure optimization, we find that among all the three structure parameters (i.e. g, d and N), g is the most important one, and generally smaller g brings about larger $f_{FEG}$ and wider high-FEG region (Fig. 2d).

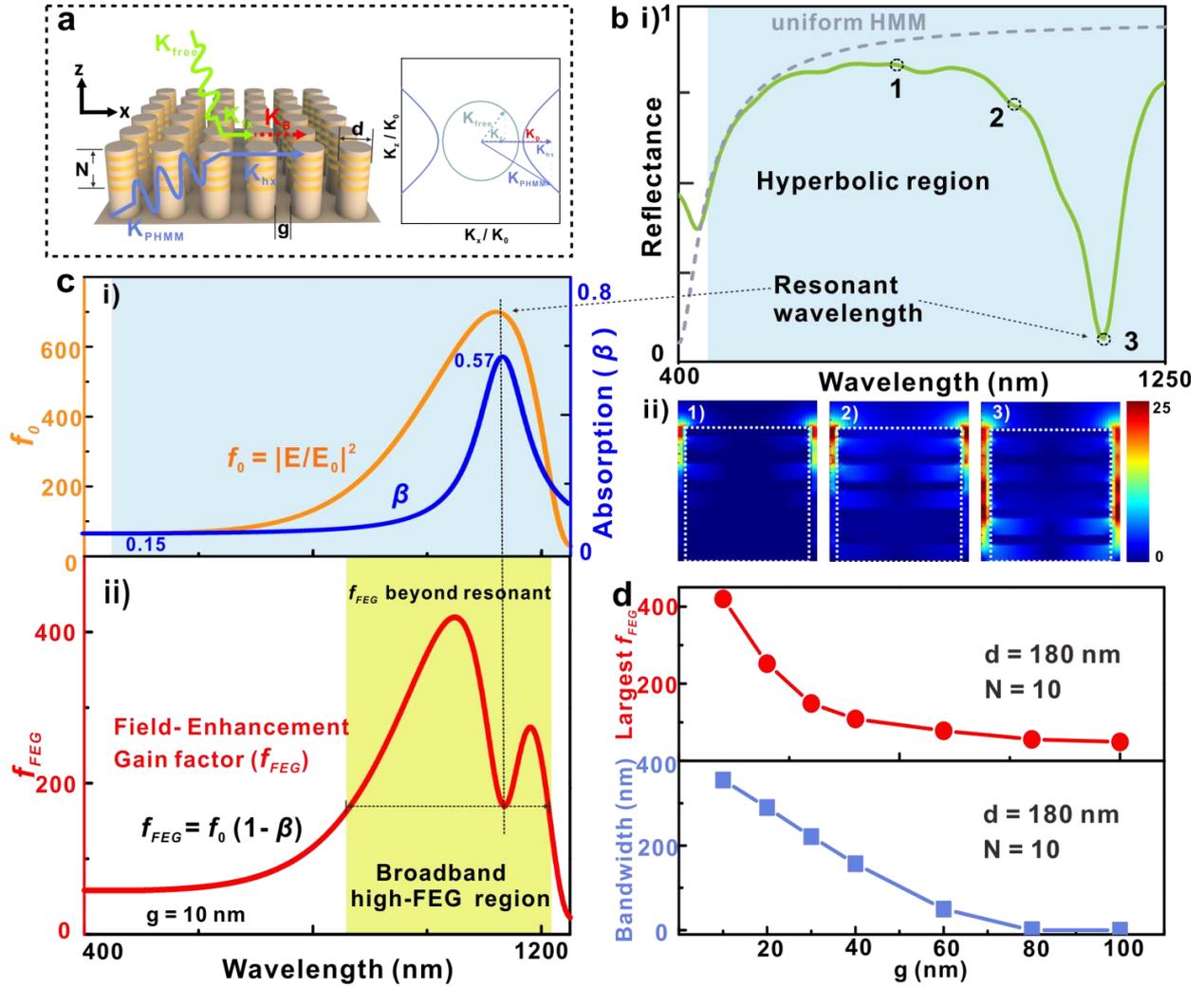

FIG. 2. Broadband high-FEG region in PHMM. (a) Schematic of the PHMM model which is arrays of nanopillars with alternating Ag and $Al_2O_3$ layers (left column). The wave-vector matching condition is shown in the right column. Normally, the wave-vector inside PHMM ($K_{PHMM}$) is larger than that in the free space ($K_{free}$). Since the Bragg scattering occurring from periodic nanopillars can generate extra wave-vector of $K_B$, wave-vector matching can be satisfied, such as $K_{hx} = K_{fs} + K_B$ ($K_{hx}$ and $K_{fs}$ are the x components of $K_{PHMM}$ and $K_{free}$, respectively). (b) i) The reflection spectra of a uniform HMM (dash line) and the PHMM (solid line). ii) The XZ plane electric field distribution of PHMM at different wavelengths marked in i). (c) i) $f_0$ and $\beta$ dependent on wavelength; ii) $f_{FEG}$ dependent on wavelength. The PHMM used here has g = 10 nm, d = 180 nm, N = 10, respectively. (d) The variation of the largest $f_{FEG}$ value and high-FEG region bandwidth with g.

Before optical measurements, we introduce the fabrication technique of our PHMM structure, because it is quite different from previous literatures. If following the existing approaches in previous studies, the fabrication should be relied on the etching of a uniform HMM by using focused ion beam (FIB) or inductively coupled plasma (ICP) [25,26]. Such methods are not suitable for our design due to their limitation of gap distance (minimal g only reaches 40~50 nm) and large roughness caused by etching processes. Therefore, we developed a semi-bottom-up method for our sample fabrication (Fig. 3a). In short, an array of dielectric nanopillars was firstly patterned by mature electron beam lithography (EBL) techniques (Fig. 3a i), which possesses higher resolution than FIB or ICP. Then, using the dielectric nanopillar array as template, five pairs of Ag (~10 nm) and $Al_2O_3$ (~20 nm) layers were alternately deposited by using electron beam (EB) evaporation (Fig. 3a ii). According to the scan electron microscopy (SEM) images (Fig. 3b and c), our PHMM were well-fabricated with uniform morphology as theoretical design. The smallest gap distance can reach as small as 20 nm, which promises large surface enhancements in the high-FEG region. Note that a uniform HMM was also formed on the substrate during our fabrication. However, such HMM hardly interacts with light from its outside because of momentum mismatching, so its existence hardly brings about influence to the PHMM (**Fig. S4**).

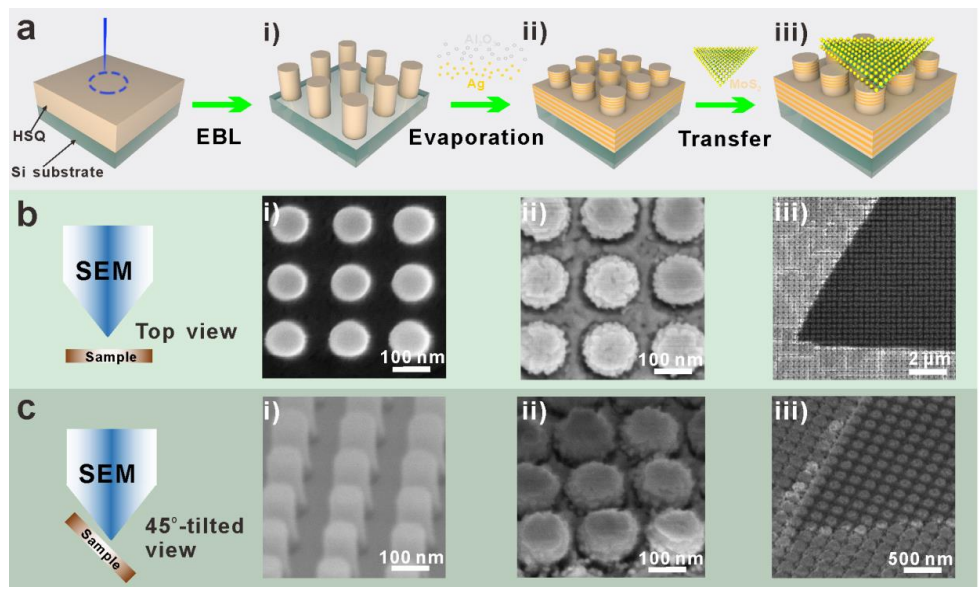

FIG. 3. Sample fabrication. (a) The fabrication scheme. i) After the coating of HSQ on a Si substrate, EBL technique is performed to fabricate dielectric nanopillar arrays; ii) then, EB evaporation is applied to deposit pairs of Ag (~10 nm) and $Al_2O_3$ (~20 nm) layers; iii) finally, a monolayer of $MoS_2$ is transferred onto the PHMM. (b) and (c) are the SEM images of our sample in each steps, respectively from top and $45^0$-tiled views.

In our nonlinear optical measurements, we obtained new records of second harmonic generation (SHG) enhancement factor and conversion efficiency of monolayer $MoS_2$ induced by the optimal FEG mode, which experimentally confirms the advantage of the optimal FEG mode and the $f_{FEG}$ factor. The reason for selecting the monolayer $MoS_2$ is because it has good optical stability, large instinct SHG response, and especially the $MoS_2$ of monolayer shows better homogeneity than few-layer ones in optical response which can help to ensure reproducibility of our measurements [27]. According to previous studies of $MoS_2$, the bandgap of monolayer $MoS_2$ c-exciton has been found locating ~430 nm [28,29], and then the SHG enhancement of $MoS_2$ should occur at ~860 nm, matching the half frequency of the c-exciton [30]. Herein, we delicately designed and fabricated two PHMM samples to make the resonant mode of the first sample and the optimal FEG mode of the second sample match with the SHG with ~860 nm, respectively shown in Fig. 4a and 4b. It is noticed that the resonant mode of the second sample is at ~960 nm. For both of the two samples, the experimental reflection spectra agree well with simulation ones.

For the first sample with the resonant mode at 860 nm, we got SHG enhancement of ~10000 times, which is 5 times of the best reported result (~2000) of monolayer $MoS_2$ when its SHG response was boosted by the gap plasmon resonant mode [15]. Moreover, we can obtain much stronger enhancement induced by the optimal FEG mode nearly at 860 nm, achieving an enhancement of 22145-fold (Fig. 4c). The corresponding SHG conversion efficiency of $MoS_2$ is calculated as high as $1.1 \times 10^{-6}$ $W^{-1}$, which is two orders of magnitude larger than previous report of monolayer $MoS_2$ [16]. These SHG enhancement records were carefully calculated, and their validities were also analyzed in theory, showing good agreement with experimental observations (**Note 1**). In addition, we observe a broadband high-FEG region over ~80 nm (820-900 nm), in which the SHG enhancements are larger than that caused by the resonant mode. The large enhancement allows us get strong SHG signal under weak excitations (lower than 1 mW, i.e. 0.051 mW/$\mu m^2$), greatly improving the stability of SHG and facilitating corresponding applications in future (**Fig. S5**). Fig. S5 also shows that the monolayer $MoS_2$ can be damaged under excitation of 10 mW with lasting time over 4 min. Such situation has been completely avoided during our optical measurement, promising the correctness of our results.

We have demonstrated the superiority of the optimal FEG mode and the FEG factor $f_{FEG}$ over the resonant mode and the FE factor $f_0$ in enhancing SHG enhancements of monolayer $MoS_2$ (detailed information about our experimental and theoretical methods are presented in **Note 2**). In order to show their generality, we have theoretically investigated the radiative power enhancement of a dipole in the PHHM. It is interesting to find that the minimum values of the radiative power enhancement and quantum efficiency always appear at the resonant mode, and their maximum values always occur in close proximity to the optimal FEG mode (**Fig. S6**). Moreover, the maximum values are 1-2 orders of magnitude larger than the minimums. These results again demonstrate that the FEG factor $f_{FEG}$ and the optimal FEG mode are more universal than the traditional FE factor $f_0$ and the resonant mode for describing the light-matter interaction, opening a new way for advanced fundamental researches and highly-efficient device applications in various fields, such as nonlinear optics, photon sources, nanolasers, and displays, etc.

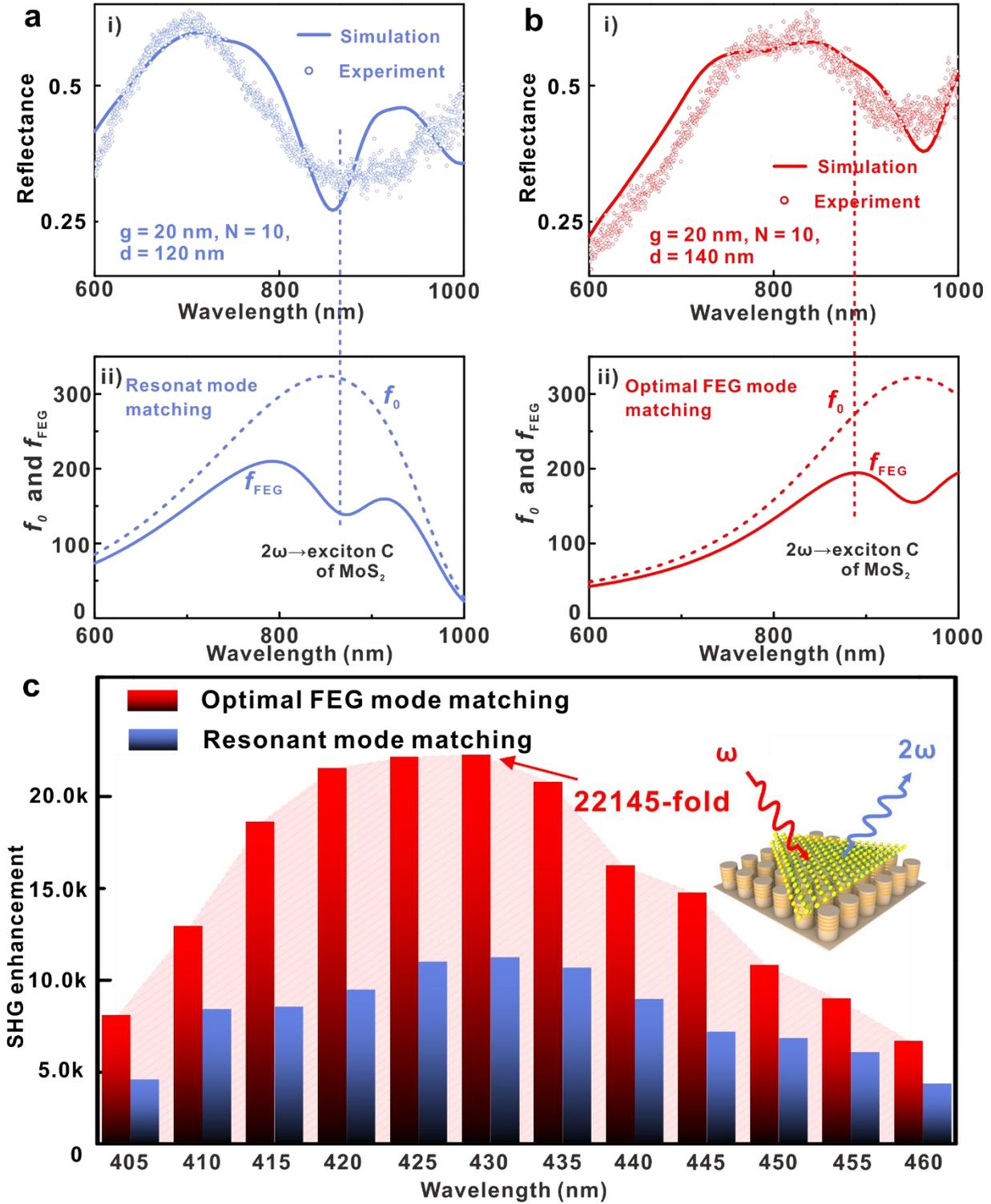

FIG. 4. Comparison of the performances of SHG enhancement enabled by the FEG and resonant modes in PHMM. (a) and (b) represent the optical behaviors of PHMM with high-k and FEG modes resonant with the half frequency of the c-exciton of monolayer MoS$_2$, respectively. In both (a) and (b), i) gives the simulation and experimental reflection spectra, and ii) shows $f_0$ and $f_{FEG}$ curves. The structural parameters for PHMMs in (a) and (b) are g = 20 nm, N =10, d = 120 nm, and g = 20 nm, N =10, d = 140 nm, respectively. (c) The SHG measurement results.

This work was supported in part by National Key R&D Program of China (2016YFA0301300), National Natural Science Foundation of China (11974437, 91750207, 11761141015), the Key R&D Program of Guangdong Province (2018B030329001), the Guangdong Special Support Program (2017TQ04C487), the Guangdong Natural Science Funds for Distinguished Young Scholars (2017B030306007), the Pearl River S&T Nova Program of Guangzhou (201806010033), the Open Fund of IPOC (BUPT) under Grant No. IPOC2019A003, and the Fundamental Research Funds for the Central Universities (20lgzd30).

*zhouzhk@mail.sysu.edu.cn，and wangxueh@mail.sysu.edu.cn.